\documentclass{iopart}

\usepackage{iopams,epsfig}

\begin{document}

\title{
Stress transmission in granular matter 
}
\author{ 
T. Aste, 
T. Di Matteo$^1$ 
and E. Galleani d'Agliano 
\footnote{e-mails: aste@fisica.unige.it; galleani@fisica.unige.it; tiziana@sa.infn.it.}}
\address{ 
INFM-Dipartimento di Fisica, via Dodecaneso 33, 16146 Genova Italy \\
$^1$ INFM-Dipartimento di Fisica ''E. R. Caianiello'', via S. Allende, 84081 Baronissi (SA) Italy. }

\centerline{To appear on \JPCM }
\centerline{\small (20/09/2001) }

\begin{abstract}
The transmission of forces through a disordered granular system is studied by means of a geometrical-topological approach that reduces the granular packing into a set of layers. 
This layered structure constitutes the skeleton through which the force chains set up. 
Given the granular packing, and the region where the force is applied, such a skeleton is uniquely defined.
Within this framework, we write an equation for the transmission of the vertical forces that can be solved recursively layer by layer.
We find that a special class of analytical solutions for this equation are L\'evi-stable distributions.
We discuss the link between criticality and fragility and we show how the disordered packing naturally induces the formation of force-chains and arches.
We point out that critical regimes, with power law distributions, are associated with the roughness of the topological layers.
Whereas, fragility is associated with local changes in the force network induced by local granular rearrangements or by changes in the applied force.
The results are compared with recent experimental observations in particulate matter and with computer simulations.
\end{abstract}




\section{Introduction}
\begin{figure}
\begin{center}
\mbox{\epsfig{file=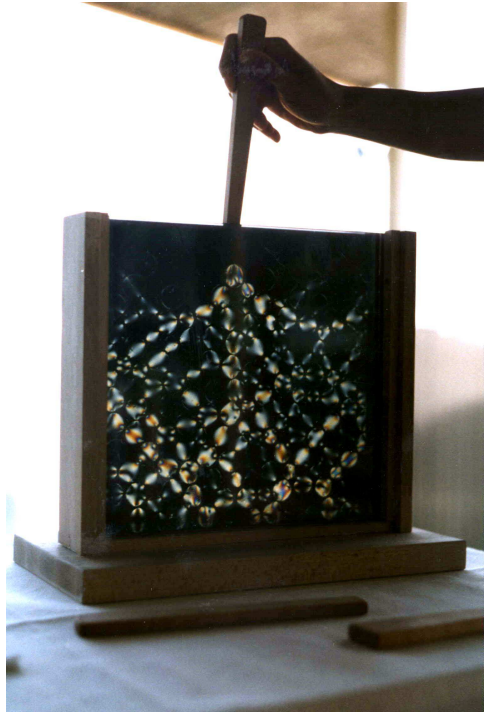,width=2.5in,angle=0}}
\end{center}
\caption{
Force chains through a disordered granular packing made by a mixture of photoelastic disks with two different sizes.
The luminosity is proportional to the amount of stress carried \cite{asteSC}.
}
\label{f.chains}
\end{figure}

In recent years many experiments \cite{Drescher,Traves,Jeager,Liu,Miller,Jeager96,Mueth,Lovoll,Blair}, simulations \cite{Radjai96,Radjai98,Tkachenko} and theoretical approaches \cite{Coppersmith,Nicodemi,Rajc,Claudin,Claudin98} have been devoted to the study of propagation of forces through granular aggregates.
One of the interesting features that rise from these studies is the formation of strong inhomogeneities in the spatial distribution of contact forces (formation of force chains), where a small fraction of the grains carries most of the total force (see Fig.\ref{f.chains}).
These chains typically extend on space scales much larger than the grain dimensions and are associated with broad stress distributions \cite{Jeager,Liu,Jeager96,Mueth,Lovoll,Radjai96,Radjai98}.
Moreover, simulations and experiments show strong rearrangements of the force-chains under small changes in the compression axes suggesting a `fragile' behavior associated with the force chain structure \cite{Radjai98,Claudin,Claudin98}.
In the present paper we show that the formation of force chains, the broadening of the force distribution and the `fragile' behavior are direct consequences of the topological disorder in the granular packing structure.

In order to study the static stress distribution, we use a scalar model where the force balance is considered in the vertical direction only and torques are neglected.
This approach is similar to several theoretical works already proposed in the literature \cite{Liu,Radjai98,Coppersmith,Nicodemi,Claudin} 
and in particular to the well known $q$ model \cite{Liu,Coppersmith}. 
In our model however, we explicitly consider a \emph{disordered} arrangements of grains (instead of placing the grains on a regular array and introducing randomness in the transmission term, as for the $q$ model).
This leads to strong differences and some important, novel consequences (see also \ref{A1}).

\section{Topological characterization of the granular packing}

\begin{figure}
\begin{center}
\mbox{\epsfig{file=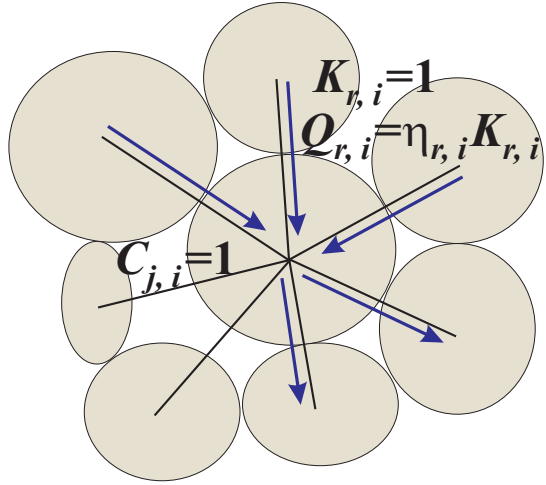,width=2.8in,angle=0}}
\end{center}
\caption{
A granular packing is topologically described by the neighboring matrix $C_{i,j}$ (lines). 
The transmission of the weight is through adjacent grains and this structure is the Dodds network which is defined by the `kissing' matrix $K_{i,j}$ (arrows). 
The force propagates in one direction and inhomogeneously among adjacent neighbours. 
This can be described by the weight transmission matrix $Q_{i,j}=\eta_{i,j}K_{i,j} $.}
\label{f.1}
\end{figure}

The complete description of a granular structure in 2 and 3 dimensions requires in general the information about the positions, sizes and shapes of all the grains. 
But, to describe the propagation of the weight (vertical forces), one can use only the topological information about the neighbors and the contacts among grains combined with the knowledge about the direction and intensity of the transmission among neighboring grains (see Fig.\ref{f.1}).

To univocally define the neighboring structure we use the `Navigation Map' \cite{Med}, which consists in the construction of space-filling cellular structure with generalized Vorono\"{\i} \cite{Voronoi} cells where the faces are made by the set of points equidistant to the surfaces of the grains. 
From this construction two grains are first neighbors if they share a common face.
The neighboring structure is topologically defined by the matrix ${\mathbf C}$:
\begin{equation}\label{Cij}
C_{i,j} =
\cases{
    1 & \mbox{if grain $i$ is first neighbor of grain $j$}, \\
    0 & \mbox{otherwise},
}
\end{equation}
which is symmetric with zeros on the diagonal.
This matrix describes the neighboring structure of the packing. 
But, by construction, two neighboring grains $i,j$ (with $C_{i,j}=1$) might not be in contact, whereas two grains which are in contact are surely neighbors.
Since the force propagates only through grains which are in contact, it follows that to study the force-propagation one must also take into account the topological structure of the inter-grain contacts (Dodds network) \cite{Dodds}.
This can be done by defining the `kissing' matrix $\mathbf{K}$:
\begin{equation}\label{Kij}
K_{i,j} =
\cases{
    1 & \mbox{if grain $i$ is in contact with grain $j$}, \\
    0 & \mbox{otherwise}.
}
\end{equation}
(with $K_{i,j}=K_{j,i}$, and $K_{i,i}=0$).
The number of grains which are in contact (`Kiss') with a given grain $i$ is
\begin{equation}\label{nni}
  n_i = \sum_{j=1}^N K_{i,j} \;\;,
\end{equation}
where $N$ is the total number of grains in the system.
An upper bound on the number of adjacent grains $n_i$, comes from geometrical considerations: it is indeed known that no more that 6 equal disks and no more than 12 equal spheres can stay in touch with a central one (kissing numbers) \cite{Conew,AsteBook}.
On the other hand, when the sizes can fluctuate and the grains can assume different shapes, these numbers might change dramatically.
But, for grain sizes distributed in a finite range, one can always find an upper bound on the number of neighbors in contact with a given grain.
For instance, when the grain sizes are homogeneous and the shapes isotropic, these upper bounds are very close to the above values of 6 and 12, in 2 and 3 dimensions respectively.
A lower bound on $n_i$ is given by the mechanical stability condition, as we will see shortly.

\vspace*{0.5cm}

\section{Weight propagation}

We study the propagation of the vertical component of the forces (weights) from grain to grain. 
This is a conservative system: the total weight on the bottom of the stack is equal to vertical force applied on the top plus the proper weights of the grains in the stack (we neglect the presence of walls that we put at the infinite).

We describe the propagation of the vertical component of the forces inside the granular structure in term of following equation:
\begin{equation} \label{W}
w_i = w_i^0 + \sum_{j=1}^N w_j Q_{j,i} \;\;\;\;,
\end{equation}
with $w_i$ being the vertical force applied at the barycenter of grain ``$i$''.
The quantities $Q_{j,i}$ are the elements of the \emph{weight transmission matrix}
$\mathbf{Q}$; they are associated with the kissing matrix through the relation
$Q_{j,i}=\eta_{j,i}K_{j,i}$,
where the quantities $\eta_{j,i}$ take values between 0 and 1.
These quantities are asymmetric: if $\eta_{j,i} > 0$ then $\eta_{i,j}=0$ (but not vice versa), giving in this way the orientation of the force network.

The symbol  $w_i^0$ indicates the proper weight of grain $i$ with an exception for the top layer of the granular stack where one can apply an external force.
In this case, the symbol  $w_i^0$ indicates the proper weight of grain $i$ plus the portion of external force acting on this grain.

Equation \ref{W} is a rather general expression: it assumes only that -as it is- the vertical force propagates only through grains which are in contact.
Equation \ref{W} can also take into account non-linear effects if one supposes that the quantities $Q_{j,i}$ are dependent on the force $w_i$. 
However, in Eq.\ref{W} the vectorial nature of the forces and the tensorial nature of the stresses have been completely neglected.
Nevertheless, we will show that the scalar description given by Eq.\ref{W} is very powerful and can take into account complex phenomena like the formation of force chains and arches or the fragility.

In Eq.\ref{W} the whole information about the granular packing is contained in the weight transmission matrix ${\mathbf Q}$ which must therefore be consistent with a real connection network in a physically-realizable granular packing.
This gives some constraints on $\mathbf Q$.
The conservation of the total weight implies that the sum over each column of the matrix $\mathbf{Q}$ must be equal to 1, i.e.
\begin{equation}\label{Wcons1}
  \sum_{j=1}^N Q_{j,i} =1 \;\;\;\;.
\end{equation}

The stability under gravity imposes that each grain must lay on at least
3 other grains in 3 dimensions or 2 grains in two dimensions, therefore
\begin{equation}\label{Wcons2}
s_i =  \sum_{j=1}^N \theta(Q_{j,i}) \ge D \;\;\;\;,
\end{equation}
with $D$ the dimensionality of the space and $\theta(x)$ the step function ($\theta(x)=1$ when $x > 0$ $\theta(x)=0$ when $x \le 0$).

The grain $i$ receives the vertical force from a number
\begin{equation}\label{mi}
m_i =  \sum_{j=1}^N \theta(Q_{j,i})  \;\;\;\;
\end{equation}
of neighbors, and shares its total weight $w_i$ through
\begin{equation}\label{si}
s_i =  \sum_{j=1}^N \theta(Q_{i,j})  \;\;\;\;
\end{equation}
neighbors.
Clearly, $m_i + s_i  = n_i$ is the number of grains adjacent to $i$ and, from the stability condition (Eq.\ref{Wcons2}), it must be: $n_i \ge D$, giving therefore a lower bound on this quantity.

\section{A linear equation}

We now seek for solutions of Eq.\ref{W}.
Let first consider the special case where the elements $Q_{i,j}$ of the weight transmission matrix depend on the granular packing structure only (i.e. the $Q_{i,j}$ don't depend on $w_j$).
In this case Eq.\ref{W} is a linear equation, and its solution can be written straightforwardly.
Indeed, an equivalent way to write Eq.\ref{W}, is:
\begin{equation}\label{Wlin}
\sum_{j=1}^N  w_j ( \delta_{j,i}-Q_{j,i} ) = w_i^0 \;\;\;\;,
\end{equation}
with $\delta_{i,j}$ the Kroneker's delta.
When the $Q_{j,i}$ are independent of $w_j$, Eq.\ref{Wlin} is a set of $N$ linear equations with $N$ unknowns.
If the determinant of $\mathbf{I-Q}$ is different from zero,
this system will admit a unique solution.
Physically this condition corresponds to impose that in the force network
there are no closed oriented rings.
In terms of the elements of the force transmission matrix this implies
that, for any given $i$, and for any arbitrarily chosen set of sites indexes $\{ j_1,...,j_k \}$, it must hold
\begin{equation}\label{Ring}
  Q_{i,j_1}  Q_{j_1,j_2}  Q_{i_2,j_3} ... Q_{j_k,i} = 0  \;\;\;\; .
\end{equation}

Note that this condition implies ${\mathbf{Q}}^x = 0$ for any $x > N$.
Indeed it is impossible to make a non-intersecting path with a length
larger than the number of grains in the system.

Equation \ref{Wlin} can be also written in a vectorial notation\begin{equation}\label{Wmat}
{ \mathbf{w}}({ \mathbf{I - Q}}) = { \mathbf{w^0}}
\end{equation}
and, when the determinant of $\mathbf{I-Q}$ is different from zero, its solution is 
\begin{equation}\label{WmatSol}
{ \mathbf{w}} = { \mathbf{w^0}}({ \mathbf{I - Q}})^{-1} \;\;\;.
\end{equation}

Although Eq.\ref{WmatSol} is formally trivial, its solution cannot be easily pursued analytically in the general case and it might become numerically untreatable for large samples.
Therefore, in the following we develop an alternative approach which considers the system as structured into a set of layers and makes possible to solve Eq.\ref{W} recursively.

\section{Construction of the layered structure}

Any granular packing can be formally reduced into a stacking of layers where the vertical force components are transmitted downward from layer to layer.   
The elements of these layers can be single grains or sets of grains that collectively behave as a single element that receives the weight from the layer above and transmits it to the layer below. 

In order to construct such a layered system, let first label all the grains at the top of the stacking (on which -eventually- the external load is applied) as belonging to the layer zero (${\mathcal{L}}_0$).
Let also consider as belonging to ${\mathcal{L}}_0$ all the grains which transmit some weight to grains already labelled as belonging to ${\mathcal{L}}_0$.
Once defined ${\mathcal{L}}_0$, the first layer (${\mathcal{L}}_1$) is made by the set of grains which are neighbors of grains in ${\mathcal{L}}_0$ and don't belong to ${\mathcal{L}}_0$. 
Moreover, are belonging to ${\mathcal{L}}_1$ all the grains which are not yet classified in ${\mathcal{L}}_0$ and that transmit some force to grains already classified in ${\mathcal{L}}_1$.
Analogously, in general, the layer ${\mathcal{L}}_\alpha$ is constituted of grains ($i$) that are neighbors of grains ($j$) in layer ${\mathcal{L}}_{\alpha-1}$ and don't belong to any already classified layer 
($i \in {\mathcal{L}}_{\alpha}$ if $i \notin {\mathcal{L}}_{\beta}$ with $\beta < \alpha$, and if exists at least one $j \in {\mathcal{L}}_{\alpha-1}$ such that $C_{i,j}=1$).  
In addition, to ${\mathcal{L}}_{\alpha}$ belong all the non-classified grains which transmit some force to grains which have been already classified in ${\mathcal{L}}_{\alpha}$ ($i \in {\mathcal{L}}_{\alpha}$ if $i \notin {\mathcal{L}}_{\beta}$ with $\beta < \alpha$, and if exists at least one $j \in {\mathcal{L}}_{\alpha}$ such that $\eta_{i,j}>0$).

In the system of layers that we have now constructed we have two kinds of grains: {\bf (i)} some grains receive the weight from the layer above only, and transmit it only to the layer below; {\bf (ii)} other grains receive or transmit weight to grains in the same layer.
We want to reduce the packing to a set of `elements' where every element $k$ (which belongs to some layer, $k \in {\mathcal{L}}_\alpha$) receives the weight from elements ($j$) in layer above ($j \in {\mathcal{L}}_{\alpha-1}$) and transmits the weight to elements $j'$ in the layer below ($j' \in {\mathcal{L}}_{\alpha+1}$).
To this purpose we identify as single \emph{elements} of the layered structure the grains of type {\bf (i)}, and the clusters of grains of type {\bf (ii)}. 

\begin{figure}
\begin{center}
\mbox{\epsfig{file=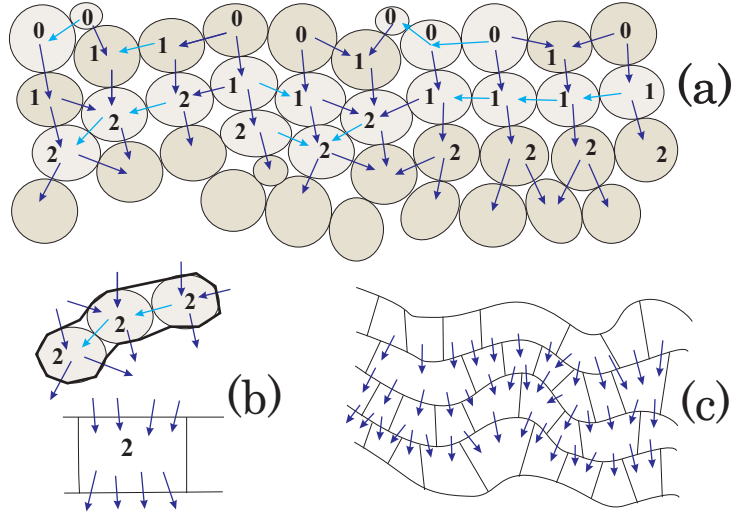,width=3.8in,angle=0}}
\end{center}
\caption{
({\bf a }) The layer ${\mathcal{L}}_{0}$ is made by the set of grains on which the force is applied plus all the grains that transmit force to them.
${\mathcal{L}}_{\alpha}$ is made of the grains which are neighbors to ${\mathcal{L}}_{\alpha-1}$ plus the grains which transmit some force to these grains in ${\mathcal{L}}_{\alpha}$ but have been not yet classified.
({\bf b }) Clusters of adjacent grains in the same layer which transmit the force among themselves are considered as single `elements'.
({\bf c}) The resulting structure is an SSI \cite{asteSSI} layered packing made of elements which receive some weight from the layer above and transmit it to elements in the layer below.
}
\label{f.layers}
\end{figure}

A schematic graphical representation of the procedure to identify the granular layers and their elements is given in Figure \ref{f.layers}.

\vspace*{0.5cm}

\section{ Recursive solutions in layered packings }

In the layered system that we have constructed in the previous section the weight is transmitted layer by layer from the top of the staking (layer ${\mathcal{L}}_0$, which receives the external load) down to the bottom layer ${\mathcal{L}}_h$ (layer at distance $h$ from the top one).

In analogy with Eq.\ref{W}, it follows that the weight on a given element in layer ${\mathcal{L}}_\alpha$ ($\alpha \ge 1$) is given in terms of the weights of the elements on the layer above ${\mathcal{L}}_{\alpha-1}$ through the following equation
\begin{equation}\label{WL-L}
w_i
=
w_i^0
+
\sum_{j \in {\mathcal{L}}_{\alpha-1}} w_{j} Q_{j,i}
\;\;\;\;\; \mbox{with $i \in{\mathcal{L}}_\alpha$.}
\end{equation}
Note that in this case the indices `$i$' and `$j$' refer to \emph{elements} of the packing instead of \emph{grains}.
The definition of $Q_{i,j}$ is identical to the one given before, but it refers to a granular structure which has been reduced to a system of elements disposed into layers which make an SSI network \cite{asteSSI} (see Fig.\ref{f.layers}c).
The forces on the top layer are known: $w_i = w^0_i$ (for $i \in {\mathcal{L}}_0$).
Therefore, by using recursively Eq.\ref{WL-L}, one can calculate step by step the forces $w_i$ from the first layer to the bottom. This recursive strategy can be applied to solve Eq.\ref{WL-L} even in the non-linear case, with the only assumption that the quantities $Q_{j,i}$ must depend only on the forces applied from the layers above.

\vspace*{0.5cm}
\begin{figure}
\begin{center}
\mbox{\epsfig{file=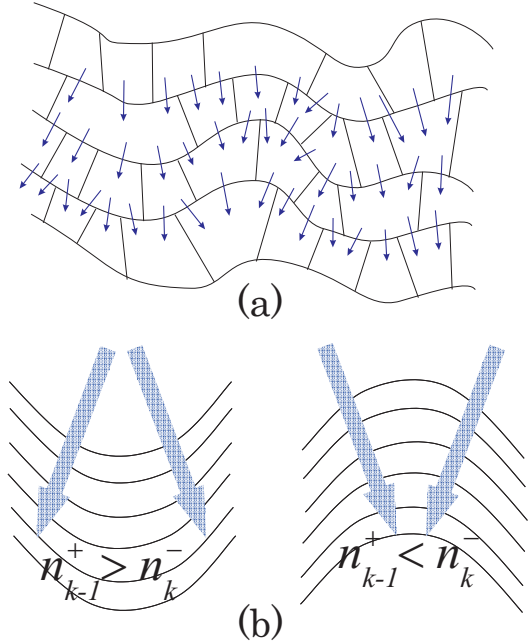,width=3.in,angle=0}}
\end{center}
\caption{
({\bf a}) The layers in disordered granular packings are made of sets of elements placed on rough surfaces (wiggled lines in two dimensions) that make dimpled landscape rich of `valleys' and `mountains'. 
For geometrical reasons, successive layers have similar local curvature and the roughening is persistent, up to a certain distance, through the layers.
 ({\bf b})  The force is concentrated (or `focalized') when transmitted through `mountains' whereas it is diffused (`defocalized') in the `valley' region.
}
\label{f.rough}
\end{figure}

\section{Force chains, arches and fragility}

In experiments and computer simulations of force propagation through granular matter an important fact is observed: only a small part of grains carries the most part of the weight.
The force propagates through `chains', and these chains generate `arches'.
The formation of these chains implies that there are some region of the packing where the weight tends to \emph{concentrate in}, and there are other regions where the weight tends to \emph{diffuse out}.

In order to understand this mechanism in term of our layered structure, let us consider the shape of these layers.
In a crystalline-ordered packing (an FCC stacking of balls, for instance) the layers are made of sets of elements disposed on flat surfaces which are parallel to each other (or straight lines in 2D packings).
When disorder is introduced, these surfaces start to bend and become rough.
The global curvature will rest zero, but locally `valleys' and `mountains' will make a dimpled landscape.
These rough surfaces are staked one upon the other and therefore the roughening of neighboring layers must be \emph{similar} and the local curvature is expected to be preserved for a few layers (see Fig.\ref{f.rough})

\begin{figure}
\begin{center}
\mbox{\epsfig{file=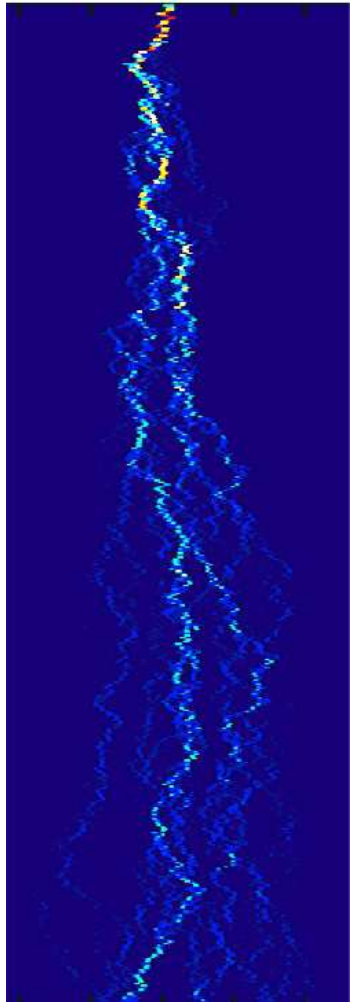,width=1.8in,angle=0}}
\end{center}
\caption{
Force chains resulting from a computer simulation of weight transmission through an SSI granular system.
}
\label{f.simul}
\end{figure}

Any given element of the layered packing receives the weight from the elements in the layer above and distributes it among the neighbors in the layer below. 
Because of the disorder, some elements have more neighbors in the layer above than others.
Clearly, if we had no structural correlation among the layers, we expect that the elements that receive the weight from more elements will result -in average- with a larger weight than the others with fewer neighbors above.
This mechanism would be amplified layer after layer if this property of abundance/deficiency of neighbors would locally propagate through the layers.
This is indeed the case in granular packings where, for geometrical reasons, the roughening of the layers must be persistent through a certain number of layers.
The persistence length is strictly related with the average chain length and gives the measure of the amplification effect.
In particular, we have that elements in `valleys' tend to dissipate the force (\emph{defocalizing effect}) by propagating it to a larger number of neighbors with respect to the number from which they receive it. 
Whereas, elements in the `mountains' tend to concentrate the weight (\emph{focalizing effect}) by transmitting the force to a smaller number of elements with respect to the number from which they receive it (see Fig.\ref{f.rough}).
The formation of arches in an `hands-on' simple experiment \cite{asteSC} made with photoelastic material between two polarizing filters, is shown in Fig.\ref{f.chains}.
The result of a computer simulation of weight propagation trough an SSI structure is shown in Fig.\ref{f.simul} \cite{AsteSimulSSI}.
As one can see, most of the stress propagates through \emph{force-chains} which make \emph{arches} by self-interconnecting into an intricate network.

\vspace*{0.5cm}

The readjustment of a few grains or a changement in the point or direction of the applied force strongly affects the layered structure.
Indeed, a local changement modifies locally the composition and curvature of the layer.
This modification propagates through the whole system by changing the roughening and the set of grains in the layers.
The final effect is a modification in the whole structure of the force-chain network.
This long-range propagation associated with the focalizing/defocalizing effects, is responsible for some of the critical effects in granular matter and it is at the origin of the fragility in these systems.

\section{Force distribution: an analytical approach }

In this section we calculate analytically, in the linear case, the probability distribution for the vertical force components.
The probability $P(w,r)$ to find a force $w$ at the site $r$ is related to the probability that the grains at sites $j$ have weights $w_j$, and that the propagation of these weights through the contact matrix $Q_{j,r}$
produces exactly $w$, which formally writes (from Eq.\ref{W}):
\begin{eqnarray}\label{Pw}
P(w,r) =
\prod_{k ,j}\int_0^1  dQ_{k,r}
\Omega({\mathbf Q })
\int      dw_j
P(w_j,j)
\delta (w -  w^0_r  -  \sum_{k' } w_{k'}Q_{k',r}), \nonumber \\
\end{eqnarray}
where $\delta(x)$ is the Dirac delta function and the quantity $\Omega({\mathbf Q })$ is the probability of a given force transmission matrix ${\mathbf Q}$.

Let us introduce the Fourier transform of $ P(w,r)$ and its inverse
\begin{eqnarray}\label{FP}
\hat P(\varphi,j) &=& {\mathcal{N}} \int dw       e^{i w \varphi}      P(w,j)       \nonumber \\
     P(w,j)       &=&               \int d\varphi e^{-i w \varphi}\hat P(\varphi,j) \;\;\;,
\end{eqnarray}with ${\mathcal{N}}$ a constant which takes into account the normalization to 1.
By using this transformation, Eq.\ref{Pw} becomes
\begin{eqnarray}\label{Pw1}
P(w,r) &=& {\mathcal{N}}
\prod_{k j  }\int_0^1  dQ_{k,r}
\Omega({\mathbf Q })
\int      dw_j
\int d\varphi_j e^{-i w_j \varphi_j}
\hat P(\varphi_j,j) \nonumber \\
&&
\int d\phi e^{- i ( w - w_r^0 - \sum_{k'  }   w_{k'}Q_{k',r}) \phi  } \;\;\;,
\end{eqnarray}
where the Dirac delta function has been written as
\begin{equation}\label{dir}
  \delta(x) = {\mathcal{N}} \int d\phi e^{-i x \phi} \;\;\;.
\end{equation}
Equation \ref{Pw1} can be re-written as:
\begin{eqnarray}\label{Pw2}
P(w,r)
&=&
{\mathcal{N}} \int d\phi e^{- i(w - w_r^0)\phi}
\prod_{ k \in {\mathcal{I}}_r   }
\int_0^1 dQ_{k,r} \Omega({\mathbf Q })
\prod_{ j \in {\mathcal{I}}_r   }
\int d\varphi_j \hat  P(\varphi_j,j)
\nonumber \\
&&
\int dw_j e^{-i w_j \varphi_j} e^{i w_{j} Q_{j ,r} \phi} \;\;\;,
\end{eqnarray}
where the integral over the $Q$'s is restricted only to the set ${\mathcal{I}}_r$ of elements which are adjacent to the element $r$  and are transmitting the force to this element (i.e. $Q_{j,r} \neq 0$).
The integration over $w_j$ yields

\begin{eqnarray}\label{Pw3} 
P(w,r) = { \mathcal{N} } 
\int d \phi  e^{-i (w - w_r^0) \phi}
\prod_{ k }
\frac{1}{ {\mathcal{N}} }
\int_0^1 dQ_{k,r} \Omega({\mathbf Q })
\prod_{ j \in { \mathcal{I} }_r   }
\hat  P(Q_{j,r}\phi,j);
 \nonumber \\
\end{eqnarray}
from Eq.\ref{FP}, its Fourier transform is
\begin{eqnarray}\label{Pw4}
\hat P(\psi,r) &=&
{\mathcal{N}}^2
\int d\phi\int dw e^{iw\psi}
e^{-i (w - w_r^0) \phi}
\nonumber \\
& &
\prod_{ k  }
\frac{1}{{\mathcal{N}}}
\int_0^1 dQ_{k,r} \Omega({\mathbf Q }) 
 \prod_{ j \in {\mathcal{I}}_r   }
\hat  P(Q_{j,r}\phi,j) \;\;\;.
\end{eqnarray}
The integration over $w$ gives finally
\begin{equation}\label{Pw5}
\hat P(\psi,r) =
{\mathcal{N}}
e^{i w_r^0 \psi}
\prod_{ k }
\frac{1}{{\mathcal{N}}}
\int_0^1 dQ_{k,r} \Omega({\mathbf Q })
\prod_{ j \in {\mathcal{I}}_r  }
 \hat  P(Q_{j,r}\psi,j) \;\;\;.
\end{equation}

If we assume that $\Omega({\mathbf Q })$ factorizes:
\begin{equation} \label{Gam}
\Omega(\{ Q \}) = \prod_{ j,k } \tilde{\Omega}_j(Q_{j,k}) \;\;\;,
\end{equation}
Eq.\ref{Pw5} reduces to
\begin{equation}\label{Pw6}
\hat P(\psi,r) = 
{\mathcal{N}}
e^{i w_r^0 \psi}
\prod_{ j \in {\mathcal{I}}_r  }
\frac{1}{{\mathcal{N}}}
\int_0^1 d\zeta_j \tilde{\Omega}_j(\zeta_j)
 \hat  P(\zeta_j\psi,j) \;\;\;.
\end{equation}

\section{L\'evi-stable force distributions in a layered-structured system}

When the force propagates downward in a layered system, an element in a layer $\alpha$ will receive the weight only from a set of elements in the layer above ($\alpha-1$) and will transmit it to a set of elements in the layer below ($\alpha+1$).
Let us suppose that the probability distribution of the force depends only on the layer number $\alpha$.  
This allows to substitute into Eq.\ref{Pw6} the quantity $\hat P(\psi,r)$ with $\hat P_{\alpha}(\psi)$ 
\begin{equation}\label{Pw7L}
\hat { P}_{\alpha} (\psi) =
{\mathcal{N}}
e^{i w_\alpha^0 \psi}
\prod_{ j \in {\mathcal{I}}_r  }
\frac{1}{{\mathcal{N}}}
\int_0^1 d\zeta_j \tilde{\Omega}_j(\zeta_j)
 \hat  {P }_{\alpha-1}(\zeta_j \psi) \;\;\;,
\end{equation}
where $ w_\alpha^0 $ is the proper weight of the elements in layer $\alpha$. 

Supposing that the load on a given element in the layer $\alpha -1$ is uniformly distributed among the $n^+_{\alpha-1}$ neighbors in the layer  $\alpha$, the term $\tilde{\Omega}_j(\zeta_j)$ becomes: $\tilde{\Omega}_j(\zeta_j)=\delta(\zeta_j - 1/n^+_{\alpha-1})$.
This simplifies Eq.\ref{Pw7L} to
\begin{equation}\label{Pw8L}
\hat { P}_{\alpha} (\psi) =
{\mathcal{N}}
e^{i w_\alpha^0 \psi}
\left[
\frac{1}{{\mathcal{N}}}
\hat  {P }_{\alpha-1}(\frac{\psi}{n^+_{\alpha-1}})\right]^{n^-_{\alpha}}
 \;\;\;,
\end{equation}
where $n^-_\alpha$ is the number of grains in layer $\alpha-1$ from which an element in layer $\alpha$ receives the force.
One can directly verify that Eq.\ref{Pw8L} is satisfied by choosing $\hat P_{\alpha} (\psi)$ in the form 
\begin{eqnarray}\label{Pw10L}
\hat P_{\alpha} (\psi)
= \exp( i\sum_{\lambda=0}^\alpha \frac{c_\lambda}{ c_\alpha } w^0_\lambda \psi )
\exp ( - \mu \sigma_\alpha |\psi|^\gamma  +  i \beta \sigma_\alpha 
\tan\left(\frac{\pi \gamma}{2}\right)
\psi |\psi|^{\gamma -1} )  \;,
\nonumber \\
\end{eqnarray}
with $\mu$ and $\gamma$ arbitrary numbers and $\beta =0$ for $\gamma =1$ or $\beta$ arbitrary for $\gamma \not=1$.
The other parameters in Eq.\ref{Pw10L} are:
\begin{equation}\label{sigma}
\sigma_\alpha = \prod_{\nu=1}^\alpha n^-_\nu (n^+_{\nu-1})^{-\gamma} \;\;\;,
\end{equation}
and 
\begin{equation}\label{PwC}
{ c_\alpha } = \prod_{\nu = 1}^{\alpha} \frac{n^+_{\nu-1}}{n^-_{\nu}} \;\;\;.
\end{equation}

Expression \ref{Pw10L} might seem complicated, but it is a broadly studied expression: it is the Fourier transform of a symmetric L\'evy-Kintchine stable distribution \cite{Levy}.
It describes a probability distribution which is peaked around the value $\sum_{\lambda=0}^\alpha c_\lambda w^0_\lambda / c_\alpha$.
It has a width proportional to $\sigma^{1/\gamma}$, and it is asymmetric with a skewness proportional to $\beta \in [-1,1]$ ($\beta$ positive distribution skewed to the right, $\beta$ negative distribution skewed to the left).
The parameter $\gamma \in (0,2]$ is called characteristic exponent.
For instance, when $\gamma =2$ the distribution is a Gaussian, whereas for $\gamma =1$ and $\beta = 0$ it is a Lorentzian.
These distributions are stable with respect to the sum of a set of stochastic variables and therefore appear naturally in the context of the Central Limit Theorem.
For $\gamma <2$ they have tails which decrease as power laws with exponents that tend to $-(\gamma +1)$ for very large deviations.
In general the moment of order $n$ becomes infinite when $n > \gamma$, it turns out therefore that the variance is only defined when $\gamma =2$, whereas the average is defined for $\gamma >1$.

From Eqs.\ref{Pw10L}, \ref{sigma} and \ref{PwC} two opposite scenarios come naturally out depending on the values of the parameters $ n^+$ and $ n^-$ through the layered structure. Let first consider the case where $ n^-_\nu > n^+_{\nu-1}$ (the elements receive the weight from a number of elements which is \emph{larger} than the number of elements to which the force is transmitted). 
From Eqs. \ref{sigma} and \ref{PwC} it follows that in this case the sum $\sum_{\lambda=0}^\alpha c_\lambda w^0_\lambda / c_\alpha$ tends to diverge with $\alpha$ and consequently the value of $\sigma$ tends to infinite. 
This is the mathematical description of the \emph{formation of a force chain} where, layer after layer the force in the chain increases (\emph{focalization effect}) with an associated broadening in the force distribution which becomes a power law with average and spread that tend to infinite.
The opposite scenario corresponds to the case when $ n^-_\nu < n^+_{\nu-1}$ (the elements receive the weight from a number of elements which is \emph{smaller} than the number of elements to which the force is transmitted).
In this case, Eqs. \ref{sigma} and \ref{PwC} yield to a force-distribution which has $\sigma \rightarrow 0$ and concentrates around the proper weight of the elements.
The elements receive zero weight from above and diffuse their weights outward from this region (\emph{defocalizing effect}).
These are the uncharged grains that stay in the regions below the arches.

\section{Conclusions}
We have presented a study of the weight propagation through a granular staking which is based on a realistic geometrical-topological description of the packing structure.
The guiding idea was to analyze the disordered packing as structured in \emph{layers} made of \emph{elements} which receive the weight from the layer above and transmit it to the layer below.
This system is an SSI packing and the equation for the vertical forces can be solved recursively.
We show that the appearance of inhomogeneous force distributions into networks of \emph{force-chains} and \emph{arches} are naturally found as a geometrical consequence of the rough shape of the layers in disordered packings.
The force is \emph{focalized} into chains or diffused out and \emph{defocalized} depending on the local curvature of the layered structure.
Strong changements in the force-network structure can be generated by small rearrangements in the packing which are amplified by the geometrical correlations through the layered system.
This leads to an intrinsic fragile behavior of these systems.
We find a class of solutions for the distribution of forces which falls in the class of L\'evy-Kintchine stable distributions.
The focalizing/defocalizing effects are analytically retrieved for these solutions and associated with the local topological properties ($n^+$, $n^-$) of the layered structure.

\bigskip
{\bf Acknowledgments:}
many thanks to Mario Nicodemi for fruitful discussions.
T.A. acknowledges many discussions with N. Rivier and C. Oguey.

\appendix

\section{Special cases: the q-model and lattice-based models} \label{A1}

In the literature several models have been presented for the force propagation in granular matter. 
Most of these models assume that the granular packing is stacked in an ordered lattice-like structure and the disorder is introduced by supposing an inhomogeneous propagation of the force among neighboring grains. 
This is -for instance- the working-framework of the well-known q-model \cite{Liu,Coppersmith}.
It might be important to point out that Equation \ref{W} becomes identical to the one of the q-model if:
\begin{itemize}
\item the equation is linear ($Q_{i,j}$ independent on $w_j$);
\item the granular packing is structured in layers;
\item the force propagates only downward from layer to layer;
\item the constraint given by Eq.\ref{Wcons2} is satisfied with the equality.
\end{itemize}
Therefore, the approach presented in this paper is applicable to the q-model, but not vice versa. 
In particular, in the framework of the q-model, the layer roughening (which is the mechanism that leads us to the formation of force chains and arches) is associated with the probability that some neighboring grains transmit no force, whereas the geometrical correlations between successive layers must be taken into account by imposing some `propagation' of these broken bonds.
These are indeed the mechanisms, proposed in the literature, which allow, within the framework of a lattice-based models (as the q-model), to obtain the formation of arches and critical force distributions \cite{Nicodemi}. 
In our approach these features come directly from the geometrical nature of the layer system.


{}



\begin{thebibliography}{}
\bibitem{AsteSimulSSI}
These simulations will be presented in larger details in a forthcoming paper.
\end{thebibliography}
\end{document}